\documentclass[final,preprint,aps,showpacs,tightenlines,fleqn]{revtex4}
\usepackage{bm}
\usepackage{dcolumn}

\begin{document} 
\preprint{PNU-NTG-02/2002} 
\title {A test of the instanton vacuum 
with low-energy theorems of the axial anomaly} 
\author{M.M. Musakhanov$^{1}$\footnote{E-mail address: 
musakhanov@nuuz.uzsci.net}
and Hyun-Chul Kim $^{2}$\footnote{E-mail address: hchkim@pusan.ac.kr}} 
\affiliation{(1)Theoretical  Physics Dept, \\ 
Uzbekistan National University,\\ 
 Tashkent 700174, Uzbekistan,\\ 
(2)Department of Physics, \\ 
Pusan National University, \\ 
609-735 Pusan, \\ 
Republic of Korea} 
\date{July 2003} 
 
\begin{abstract} 
We revisit the QCD$+$QED axial anomaly low-energy 
theorems which give an exact relation between the matrix elements 
of the gluon and photon parts of the axial anomaly operator equation 
within the framework of the {\em effective action} derived from 
the instanton vacuum.  The matrix elements between the vacuum and two 
photon states and between the vacuum and two gluon states are 
investigated for arbitrary $N_f $ in the chiral limit. 
Having gauged the effective action properly, we show that the model 
does exactly satisfy the low-energy theorems. 
\vspace{1cm} 
 
\end{abstract} 
\maketitle 
%\section { Introduction} 
{\bf 1.} The quantum fluctuations may destroy the symmetries of the initial 
classical Lagrangian.  One of the most important examples is the
famous axial anomaly in gauge theories.  The axial anomaly leads to
many interesting nonperturbative features intimately related to the
topologically nontrivial structure of the vacuum.  
 
The axial anomaly in the divergence of the singlet axial-vector current in 
QCD $+$ QED brings forth a low-energy theorem for the matrix elements 
of this operator equation between the vacuum and two--photon states 
(see, for example, a review~\cite{Shifman:1988zk}): 
\begin{equation} 
\left \langle 0\left|N_f \frac{g^2}{32\pi^2}G\tilde G 
\right| 2\gamma \right\rangle = 
 \frac{N_c}{8\pi^2} \sum_{f}{e^{2}_{f}} F^{(1)}\tilde F^{(2)} 
\label{theorem} 
\end{equation} 
at $q^2 \ll m_{\eta'}^2$ ($m_{\eta'}$ is the $\eta'$-meson mass around 
1 GeV).  $N_f$ is the number of flavors, $g$ is the QCD coupling
constant with $G\tilde G= \frac{1}{2}\epsilon^{\mu\nu\lambda\sigma} 
G^{a}_{\mu\nu}  G^{a}_{\lambda\sigma}$ and $N_c$ is the number of 
colors.  Here, $G^{a}_{\mu\nu}$ stands for the operator of the gluon 
field strengths.  $e_f$ represent the electric charges of quarks.  The   
photon part of Eq.(\ref{theorem}) can be expressed explicitly as 
$
F^{(i)}_{\mu\nu} \;=\; \epsilon^{(i)}_\mu q_{i\nu} - 
\epsilon^{(i)}_\nu q_{i\mu}, 
$
where $\epsilon^{(1,2)}_\mu$ and $q_{1,2}$ are, respectively,
polarizations and momenta of photons, and $q=q_{1} + q_{2}$.  In 
perturbation theory, it leads to at least results of order 
$e_{f}^2 g^{4}$ for the left-hand side of Eq.~(\ref{theorem}). 
 
Another nontrivial low-energy theorem concerns the matrix element 
between the vacuum and two-gluon states: 
\begin{equation} 
 \langle 0| g^2 G\tilde G | 2 \mbox{gluons} \rangle = 0 
 \label{theorem1} 
\end{equation} 
at the same limit as in Eq.(\ref{theorem}).  Hence, the solution of 
these theorems given by Eqs.(\ref{theorem},\ref{theorem1}) is 
only pertinent to the nonperturbative phenomena related to the 
topologically nontrivial structure of the QCD vacuum. 
 
Without any doubt instantons represent a very important topologically 
nontrivial component of the QCD vacuum. 
The most important parameters of the QCD instanton vacuum are 
the average instanton size $\rho$ and inter-instanton distance $R$. 
Shuryak \cite{Shuryak:1981ff,Shuryak:dp} estimated these two 
parameters phenomenologically as 
$
\rho \,\simeq\, 1/3 \, {\rm fm},\, \, \, \, R\,\simeq \, 1 \, {\rm fm}. 
\label{rhoR} 
$
It was confirmed by theoretical variational calculations 
\cite{Diakonov:1983hh,Diakonov:1985eg} and recent lattice simulations 
of the QCD vacuum~\cite{DeGrand:2001tm}.  In particular, the 
spontaneous breakdown of chiral symmetry is realized very well via the 
instanton liquid model.  Hence, instantons play a pivotal and more 
significant  role in describing the lightest hadrons and their interactions, 
compared with confinement forces~\cite{Diakonov:2002mb,Chu:vi}. 
 
Some years ago, one of the authors showed that the low-energy QCD 
effective chiral action from the instanton vacuum 
~\cite{Diakonov:1985eg,Diakonov:1995qy} satisfies the low energy 
theorems with an accuracy of about 17 $\%$~\cite{Musakhanov:1996qf,Salvo:1997nf}. 
This discrepancy is due to the fact that the vector and axial-vector 
currents are not conserved because of the momentum-dependent 
dynamical quark mass in the effective 
action~\cite{Ball:1993ak,Bowler:ir,Plant:1997jr,Broniowski:1999bz}. 
The nonconservation of the N\"other currents breaks various Ward 
identities in the model.  In order to remedy these problems, 
we shall gauge the low-energy QCD effective partition function properly, 
from which the conserved currents can be derived. 
 
In the present paper, we shall continue to test the low-energy 
QCD effective chiral action from the instanton vacuum in the chiral 
limit aiming at resolving the discrepancy existing in 
Ref.~\cite{Musakhanov:1996qf}.  Having gauged the low-energy QCD effective 
partition function, we shall show that it does exactly satisfy these 
theorems as it should be. 
 
%\section{Low-energy QCD effective chiral action in the chiral limit} 
{\bf 2.} 
We have to calculate the quark determinant in the presence of 
instantons and external electromagnetic field $v_\mu$, and then average it
over the collective coordinates of the instantons to find the
effective low-energy QCD partition function.  The first thing we have
to do is to find the extended zero-modes  $\tilde\Phi_{0}$ 
for the quark placed into the instanton field $A_{\mu}^{I}$ and the
electromagnetic field $v_\mu$, which obey the following equation:
\begin{equation}
(i\rlap{/}{\partial}_x + g \rlap{/}{A}^I(x) + e
\rlap{/}{v}(x))\tilde\Phi_{0}(x) = 0   
\label{ed}
\end{equation}
The extended zero-modes satisfy the U(1) gauge transformation
\begin{equation}
\tilde\Phi_{0} = L \Phi_{0}' ,\,\,\,
L(x) =\exp \left(ie x_\mu \int^{1}_{0} v_\mu (\alpha x)d\alpha\right),
\label{ezm}
\end{equation}
where $L$ is simply a path-ordered exponent
$P\exp(i\int_{0}^{x} v_\mu (s)ds_\mu)$ with a straight path connecting
the points $0$ to $x$ ~\cite{chretien54,Radyushkin:1998du}. 
Then Eq.~(\ref{ed}) is reduced to
\begin{equation}
(i\rlap{/}{\partial}_x + g \rlap{/}{A}^I(x) + e \rlap{/}{v}'(x))\Phi_{0}' (x) = 0,
\label{ed1}
\end{equation}
where
\begin{equation}
v_{\mu}' (x)= x_\rho \int^{1}_{0} F_{\rho\mu} (\alpha x)\alpha d\alpha
\end{equation}
has an explicit gauge-invariant form.  An approximate solution of
Eq.~(\ref{ed1}) is 
\begin{equation}
\Phi_{0}' = \Phi_0 - S^I e \rlap{/}{v}' \Phi_0 ,
\label{phi'}  
\end{equation}
which can be expanded to any desired order of $ev_\mu'$.
Here, $S^I =(i\rlap{/}{\partial} + g \rlap{/}{A}^I +im)^{-1}$ is a quark propagator 
(with a small mass $m$) in the field of a single instanton:
\begin{equation}
S^I = 
\frac{|\Phi_0\rangle\langle\Phi_0|}{im} +
S_{NZ}, \,\,\,\, S_{NZ}=
\sum_{n\neq 0 }\frac{|\Phi_n\rangle\langle\Phi_n|}{\lambda_n +im}  
\end{equation}
where $\Phi_n$ is defined by $(i\rlap{/}{\partial} + g\rlap{/}{A}^I
)|\Phi_n \rangle =\lambda_n  |\Phi_n \rangle$ 
(with the convention $\lambda_0 =0$).  The quark propagator of the
non-zero mode $S_{NZ}$ in the single instanton field was derived in 
an exactly closed form~\cite{BCCL78}.  It is clear from this formula that 
the quark propagator is reduced to the free one at a short distance as well
as a long distance.  We follow here the argument by
Ref.~\cite{Diakonov:1985eg} and will assume the following relation:   
\begin{equation}
S_{NZ}(x,y)\simeq S_{0}(x-y).
\label{eq:snz}
\end{equation}
Combining Eq.(\ref{eq:snz}) with Eqs.(\ref{ezm},\ref{ed1},\ref{phi'}),
we obtain the extended zero mode
$
\tilde\Phi_{0} = L \Phi_{0}' .
$
Now, we consider the quark propagator $\tilde S$ in the presence of
the single instanton $A_{\mu}^{I}$ and $v_\mu$ with the assumption
given by Eq.(\ref{eq:snz}), so that we obtain for the quark propagator
$\tilde S^I$: 
\begin{eqnarray}
\nonumber
\tilde S^I &=& (i\rlap{/}{\partial} +g\rlap{/}{A}^I + e\rlap{/}{v}
+im)^{-1} = S^I - S^I e\rlap{/}{v}S^I +\cdots  
\\
\nonumber
&=&
\frac{|\Phi_{0}\rangle\langle\Phi_{0}|}{im} + S_{NZ} -
\left(\frac{|\Phi_{0}\rangle\langle\Phi_{0}|}{im} + 
S_{NZ}\right)e\hat v\left(\frac{|\Phi_{0}\rangle\langle\Phi_{0}|}{im}
+ S_{NZ}\right)+\cdots  
\\
\label{S^I}
&\simeq& \tilde S_{0} +
\frac{|\tilde\Phi_{0}\rangle\langle\tilde\Phi_{0}|}{im}, 
\end{eqnarray}
where $\tilde S_{0} = (i\rlap{/}{\partial} +e\rlap{/}{v} +im)^{-1}$.

Now, we are in a position to consider the ensemble of instantons
$
A_\mu = \sum_I A_{I \mu},
$
representing the instanton-liquid picture of the QCD vacuum.
This sum consists of both instantons and antiinstantons.  The quarks
are moving in the presence of the instanton ensemble $A_\mu$ and 
external electromagnetic field $v$.  Thus, we can derive an extended
zero-mode approximation for the quark determinant.   

The quark propagator $\tilde S$ can be expanded with respect to a
single instanton:
\begin{eqnarray}
\label{prop_sum} 
 \tilde S &=&\tilde S_{0} - \sum_I\tilde S_{0} g \rlap{/}{A}_I\tilde S_{0}
 + \sum_{I,J}\tilde S_{0} g\rlap{/}{A}_I
\tilde  S_{0} g\rlap{/}{A}_J\tilde S_{0} + \ldots \\
\label{prop_hop}
 &=&\tilde S_{0} + \sum_I (\tilde S_{I} -\tilde S_{0}) + \sum_{I\neq J} 
 (\tilde S_{I} -\tilde S_{0})\tilde S_{0}^{-1} (\tilde S_{J} -\tilde S_{0}) \\
 & & + \sum_{I\neq J,\,J\neq K}  (\tilde S_{I} -\tilde S_{0})
\tilde S_{0}^{-1} 
 (\tilde S_{J} -\tilde S_{0})\tilde S_{0}^{-1}(\tilde S_K-\tilde S_{0}) 
+ \ldots \; .\nonumber  
\end{eqnarray}
Here, the indices $I,J,K,\cdots$ designate both instantons and anti-instantons. 
In Eq.(\ref{prop_hop}), we have resummed the contributions corresponding 
to an individual instanton. $\tilde S_{I}$ stands for the quark
propagator in the fields of the single instanton $I$ as well as in the
external electromagnetic field $v$ and represents the sum of the zero
and non-zero modes.  At a large distance from the center of the
instanton, $\tilde S_{I}$ becomes the propagator $\tilde S_{0}$ in 
the presence of the external field $v$.  Equation~(\ref{prop_hop}) is 
obviously an extended quark propagator of that from
Ref.~\cite{Diakonov:1985eg} (see also review~\cite{Schafer:1996wv}). 
Thus, Eq.~(\ref{S^I}) can be written as 
\begin{equation}
\label{zm_dom}
 \left(\tilde S_{I}-\tilde S_{0}\right)(x,y) \simeq 
   \frac{\tilde\Phi_{I,0}(x)\tilde\Phi_{I,0}^{\dagger} (y)}{im}.
\end{equation}
In the present case, the expansion given by Eq.(\ref{prop_hop})
becomes 
\begin{eqnarray}
\tilde  S (x,y) &\simeq&\tilde S_{0}(x,y) + \sum_I\frac{\tilde\Phi_{I,0}(x)
  \tilde\Phi_{I,0}^{\dagger } (y)}{im}
\\
\nonumber   
&+& \sum_{I\neq J} \frac{\tilde\Phi_{I,0}(x)}{im}
      \left( \int d^4r\,\tilde\Phi_{I,0}^{\dagger}(r)
      (i\rlap{/}{\partial} +e \rlap{/}{v} +im)\tilde\Phi_{J,0}(r) \right)
     \frac{\tilde\Phi_{J,0}^{\dagger}(y)}{im} + \cdots .
\end{eqnarray}
Defining the following overlapping integrals 
\begin{equation}
 \tilde a_{IJ} =- \int d^4r\,\tilde\Phi_{I,0}^{\dagger}(r)
(i\rlap{/}{\partial} +e \rlap{/}{v} )\tilde\Phi_{J,0}(r),
\label{tildea}
\end{equation}
we are led to the expression:
\begin{eqnarray}
\label{prop_zmztilde}
\tilde S(x,y) \simeq\tilde S_{0}(x,y) + \sum_{I,J}\tilde\Phi_{I,0}(x)
    \left(\frac{1}{\tilde a + im }\right)_{IJ}
     \tilde\Phi_{J,0}^{\dagger}(y).
\end{eqnarray}

The total quark determinant should be now splitted into two parts,
{\em i.e.} low and high frequencies (with respect to a mass  
parameter $M_1 \sim \rho^{-1}$): ${\rm Det} =
{\rm Det}_{\rm low}\times{\rm Det}_{\rm high}$~\cite{Diakonov:1985eg}. 
The high-frequency part ${\rm Det}_{\rm high}$ can be written as a
product of the determinants in the field of individual instantons,
while the low-frequency one ${\rm Det}_{\rm low}$ has to be treated
approximately, would-be zero modes being taken into account only.
Thus, we get for the low-frequency part of the quark determinant as
follows: 
\begin{eqnarray}
\ln{\rm Det}_{\rm low} =\sum_f \int^{m_f}_{M_1} idm 
[{\rm Tr} (\tilde S - \tilde S_{0} ) - {\rm Tr} (\tilde S_{0}-S_0 )],
\label{detlow}
\end{eqnarray}
where
\begin{equation}
{\rm Tr} (\tilde S - \tilde S_{0} )={\rm Tr} \frac{1}{\tilde a +im}.
\end{equation}
The second term in Eq.(\ref{detlow}) can be included in the normalization
factor.   Taking into account the external gauge field
$v$, we get 
\begin{equation}
{\rm Det}_{\rm low}={\det}_{N,v} = \det \tilde B, \,\,\,\,
\tilde B_{IJ}=\tilde a_{JI} +im\delta_{JI}.  
\label{B}  
\end{equation}
If we had turned off the external field $v$, we would have obtained
exactly the same quark determinant $\det_N$ by Lee and
Bardeen~\cite{Lee:sm}.  Having  fermionized the representation of
Eq.~(\ref{B})~\cite{Musakhanov:1996qf,Salvo:1997nf}, we  
derive the gauged quark determinant in terms of the 
constituent quarks $\psi_{f}$: 
\begin{eqnarray} 
{\rm det}_{N,v} &=& \int D\psi D\psi^{\dagger} \exp(\int d^4 x 
\sum_{f}\psi_{f}^{\dagger}(i\rlap{/}{\partial}+ e_f \rlap{/}{v}) \psi_{f}) \cr 
&\times& \prod_{f}(\prod_{+}^{N_{+}}(im_{f} - \tilde{V}_{+}[\psi_{f}^{\dagger} 
,\psi_{f}]) \prod_{-}^{N_{-}}(im_{f} - \tilde{V}_{-}[\psi_{f}^{\dagger} , 
\psi_{f}])), 
\label{part-func} 
\end{eqnarray} 
where 
\begin{eqnarray} 
\tilde{V}_{\pm}[\psi_{f}^{\dagger} ,\psi_{f}]
&=&\int d^4 x (\psi_{f}^{\dagger} (x) (i\rlap{/}{\partial} + e_f
\rlap{/}{v}) \tilde\Phi_{\pm , 0} (x; \xi_{\pm}))\int d^4 y 
(\tilde\Phi_{\pm , 0} ^\dagger (y; \xi_{\pm} ) 
(i\rlap{/}{\partial}+ e_f \rlap{/}{v}\psi_{f} (y))
\label{Eq:gfm}\\
&=& 
\int d^4 x (\psi_{f}^{\dagger} (x) L_f(x)i\rlap{/}{\partial} 
\Phi_{\pm , 0} (x; \xi_{\pm}))\int d^4 y 
(\Phi_{\pm , 0} ^\dagger (y; \xi_{\pm} ) 
i\rlap{/}{\partial}L_f^{\dagger} (y)\psi_{f} (y)).
\label{Eq:gfm1} 
\end{eqnarray}
Here, we have used Eqs.(\ref{phi'},\ref{eq:snz}) to derive
Eq.(\ref{Eq:gfm1}) from Eq.(\ref{Eq:gfm}).  The range of the
integration in Eq.(\ref{Eq:gfm}) is cut at $\rho$, since it is defined
by zero-mode functions $ \Phi_{\pm , 0}$. 
In the present case, we assume the soft-photon external field $q\rho
\ll 1$ ($q$ is a photon momentum) which simplifies the gauge factor:
\begin{equation}
L_f(x) = \exp\left(ie_f(x-z)_\mu \int_0^{1}
  v_\mu(z+\alpha(x-z))d\alpha\right)   
\simeq \exp\left(ie_f(x-z)_\mu v_\mu(z)\right)
\end{equation}
Note that external $v_\mu$ field gauges not only the kinetic term of 
the effective action but also its interaction term 
$\tilde V_{\pm}[\psi_{f}^{\dagger} ,\psi_{f}]$ in 
Eq.~(\ref{Eq:gfm}).  The reason is obvious: It is the nonlocal
interaction induced by instantons.  The external $v_\mu$-field is
present here due to the factor $L_f$ attached to each fermionic line.
This factor provides us an approximate gauge invariance of the
interaction term $\tilde V_{\pm}[\psi_{f}^{\dagger} ,\psi_{f}]$ under
the gauge transformation which is valid in the soft-photon limit
$q\rho \ll 1$.  The remaining problem is to average the quark
determinant over collective coordinates $\xi_\pm$.  It is a rather
simple procedure, since the low density of the instanton medium ($\pi^2
\left(\frac{\rho}{R}\right)^4\sim 0.1$) allows us to average over
positions and orientations of the instantons independently.  In the
following, we will assume the chiral limit.  Thus,
Eq.(\ref{part-func}) gives us the gauged partition function: 
\begin{equation} 
 Z_N[v] = \int D\psi D\psi^\dagger \exp  \left(\int d^4 x \, 
 \sum_{f}\psi_{f}^{\dagger} 
(i \rlap{/}{\partial} + e_f \rlap{/}{v}) \psi_{f} \right) 
\,  \tilde{W}_{+}^{N_+}  \, \tilde{W}_{-}^{N_-}, 
\label{Z_NW} 
\end{equation} 
where 
\begin{equation} 
\tilde{W}_\pm =  \int d \xi_{\pm}\prod_{f} 
(\tilde{V}_{\pm}[\psi_{f}^{\dagger} \,\psi_{f}] ) 
= (i)^{N_{f}}\left(  \frac{4\pi^2 \rho^2}{N_c} \right)^{N_f} 
\int \frac{d^4 z}{V} \det_{f}i \tilde{J}_\pm (z) 
\end{equation} 
and 
\begin{equation} 
\tilde{J}_\pm (z)_{fg} \;=\; \int \frac {d^4 kd^4 l}{(2\pi )^8 } 
\exp ( -i(k - l)z) \, F((k+e_fv(z))^2) F((l+e_gv(z))^2) \, 
\psi^\dagger_f (k) \frac12 (1 \pm \gamma_5 ) \psi_g (l) . 
\label{J_pm} 
\end{equation} 
The form factor $F(k^2)$ is related to the 
zero-mode wave function in momentum space $\Phi_\pm (k; \xi_{\pm}) $ 
and turns out to be 
$
F(k^2) = - t \frac{d}{dt} \left[ I_0 (t) K_0 (t) - I_1 (t) K_1 (t) 
\right], \,\, t =\frac{1}{2} \sqrt{k^2} \rho, 
$
where $I_i$ and $K_i$ are modified Bessel functions.  Thus, we find 
that the gauged partition function in the soft-photon limit 
is nothing but the original partition function with the gauged form factor 
$F((k+e_fv(z))^2)$.  Details of how to derive the effective action
from the partition function are well described in 
Ref.~\cite{Diakonov:1985eg}.  The self-consistency condition of the 
exponentiation and bosonization at the common saddle-point leads to
the following equation for the dynamical quark mass $M$:   
\begin{equation} 
4 N_c \int \frac{d^4 k}{(2\pi )^4} \frac{M^2 F^4 (k)}{M^2 F^4 (k) + 
k^2} =  \frac{N}{V}, 
\label{selfconsist} 
\end{equation} 
where $N/V$ denotes the finite density of instantons and 
antiinstantons in Euclidean space. 

In the quasiclassical (saddle-point) approximation, any gluon operator 
receives its main contribution from the instanton background. 
As an example, for one instanton (anti-instanton)$I (\bar I)$, 
the following gluonic operators can be expressed in terms of the 
instanton background: 
\begin{eqnarray} 
{g^2}G^2 (x) = 
\frac{192 \rho^4}{\left[ \rho^2 + (x - z)^2 \right]^4} = f(x-z), 
\,\,\,\,\, 
{g^2}G\tilde G(x)  = \pm f(x-z) . 
\label{GtildeG} 
\end{eqnarray} 
The derivation of the correlators with the gluonic 
operator $G\tilde G$ comes down to that of the 
following partition function: 
\begin{equation} 
\hat Z_{N}  [v,\kappa] = 
Z_{N}^{-1} \int  D\psi D\psi^\dagger \exp  (- \hat S_{\rm eff}  ) , 
\label{hatZ_N} 
\end{equation} 
where the effective  action, $\hat S_{\rm eff}$ in the presence of 
an external fields $v_\mu(x)$ and  $\kappa (x)$ coupled to
$G\tilde{G}$, is given by  
\begin{eqnarray} 
-\hat S_{\rm eff}  &=& 
 \int d^4 x \, \sum_{f}\psi_{f}^{\dagger} 
(i \rlap{/}{\partial} + e_f \rlap{/}{v}) \psi_{f} 
+\left(\frac{2V}{N}\right)^{N_f - 1} 
\int d^4z \,  \det (i  M_{+} (z) \tilde{J}_{+} (z)) \cr 
&& \hspace{47pt} + 
\left(\frac{2V}{N}\right)^{N_f - 1} 
\int d^4z \, \det (i  M_{-} (z) \tilde{J}_{-} (z)) , 
\label{hatS_ef3} 
\end{eqnarray} 
where 
$
M_{\pm} (z) = 
 \left( 1 \pm \int dx \kappa (x) f(x-z)\right)^{(N_f - 1)^{-1} }M 
$ ~\cite{Musakhanov:1996qf}.
Using the bozonization again, one can show that the effective
action  $\hat S_{\rm eff}[ {\cal M}_{\pm}, v,\kappa ]$
turns out to be  
\begin{eqnarray} 
-\hat S_{\rm eff}[{\cal M}_{\pm},v,\kappa] &=& \int d^4z\left(-(N_f -
  1)  \left(\frac{2V}{N}\right)^{ - 1} 
(\det{\cal M}_{\pm} )^{\frac{1}{N_f - 1}}\right) \cr 
&+& 
{\rm Tr} \ln \left[ \rlap{/}{P} + i M F(P^2) {\cal M}_{+}F(P^2) \left( 1 + 
( \kappa  f)\right)^{N_{f}^{-1}} \frac {1}{2}(1+\gamma_5 )\right.  \cr 
&& + \left.i M F(P^2) {\cal M}_{-}F(P^2) 
\left( 1 - (\kappa f)\right)^{N_{f}^{-1}} \frac {1}{2}(1- \gamma_5 )\right] 
\label{hat S_ef5} 
\end{eqnarray} 
in the presence of the external fields $v_\mu$ and $\kappa$. 
Here, $P_\mu$ denotes the covariant momentum 
$
P_\mu := p_\mu  + e Q v_\mu 
$
with $e Q_{fg}=e_f \delta_{fg}$.  The matrices ${\cal M}_\pm$ describe 
the meson fields and $(\kappa  f) =  \int d^4x \kappa (x) 
f(x-z)$.  
 
%%%%%%%%%%%%%%%%%%%%%%%%%%%%%%%%%%%%%%%%%%%%%%%%%%%%% 
%\subsection { The low-energy theorems for the matrix 
%element between the vacuum and two-photons state} 
%%%%%%%%%%%%%%%%%%%%%%%%%%%%%%%%%%%%%%%%%%%%%%%%%%%%% 
Since the matrix element between the vacuum and two-photon state is 
defined by the  functional derivative 
$
\left. \frac {\delta \hat Z_{N}  [\kappa , v] } 
{\delta \kappa (x) \delta v_\mu (x_1) \delta v_\nu (x_2) } 
\right|_{ \kappa , v = 0}, 
$
we need to keep only the terms of order ${\cal O}(\kappa)$ and ${\cal O}(v^2)$ in 
$\hat S_{\rm eff}[ {\cal M}_{\pm}=1, v, \kappa]$ defined by
Eq.(\ref{hat S_ef5}). 
 
We will employ for the explicit calculation the Schwinger method~\cite{Schw}, 
which was developed for quantum electrodynamics and later extended to 
QCD~\cite{Vainshtein:xd},  
in order to expand $\hat S_{\rm eff}$ with respect to derivatives of  
the photon background field.  Setting ${\cal M}_{\pm}=1$, considering
the first order of $\kappa$, and defining $M(P)=MF^2 (P)$, we get 
\begin{eqnarray} 
 -\hat S_{\rm eff}[ {\cal M}_\pm=1, v, \kappa ]
={\rm Tr} \ln [ \rlap{/}{P} + i M (P)] + 
\frac{i}{N_f} {\rm Tr}\left[(\rlap{/}{P} + i M (P))^{-1}M (P)(\kappa 
  f)\gamma_5\right], 
\label{hat S_ef6} 
\end{eqnarray} 
where ${\rm Tr} = {\rm tr}_{cf\gamma}\int d^4 x\langle x|..... |x\rangle$ and 
${\rm tr}_{cf\gamma}$ denotes the trace over color, flavor, and Dirac 
spaces.  The quark propagator with the covariant derivative can be 
treated as follows: 
\begin{eqnarray} 
( \rlap{/}{P} + i M (P))^{-1}
= \left( P^2 +  M^2 (P) + \frac{eQ}{2}\sigma \cdot F 
+i[ \rlap{/}{P},M(P)]\right)^{-1} (\rlap{/}{P} - i M (P)),
\label{paropagator} 
\end{eqnarray} 
where $\sigma\cdot F=\sigma_{\mu\nu} F_{\mu\nu}$. 
We need at least four $\gamma$ matrices with one $\gamma_5$ so 
that we may obtain a nonvanishing result with the trace over Dirac space 
(${\rm tr}_\gamma$).  Thus, the relevant nonvanishing part of the 
effective action is: 
\begin{eqnarray} 
-S_{kf} &=&
\frac{1}{N_f} {\rm Tr}\left\{ \left[( P^2 +  M^2 (P))^{-1} 
\left(\frac{eQ}{2}\sigma\cdot F 
+i[ \rlap{/}{P},M(P)]\right)\right]^2\right. \cr 
&\times&\left. 
( P^2 +  M^2 (P))^{-1}(M^2(P) + i \rlap{/}{P} M(P)) 
(\kappa f)\gamma_5 \right\}. 
\label{S_kf} 
\end{eqnarray} 
We observe that $[ \rlap{/}{P},M(P )]$ is of order ${\cal O}(e)$ and 
$M$ is a function of $P^2$.  Then 
\begin{eqnarray} 
-S_{kf} &=& \frac{1}{N_f} {\rm Tr} \left[( p^2 +  M^2 (p^2))^{-1} 
\left(\frac{eQ}{2}\sigma\cdot F\right)^2 
( p^2 +  M^2 (p^2 ))^{-1} M^2(p^2 )(\kappa f)\gamma_5\right. \cr 
&-&\left\{ ( p^2 +  M^2 (p^2 ))^{-1}\frac{eQ}{2}\sigma\cdot F,\,\, 
 ( p^2 +  M^2 (p^2 ))^{-1} [ \rlap{/}{P},M(P^2 )]\right\}\cr 
&\times& 
( p^2 +  M^2 (p^2 ))^{-1} \rlap{/}{p} M(p^2 ) (kf)\gamma_5 \Big] 
 \,+\, {\cal O}(e^3 ). 
\label{S_kf1} 
\end{eqnarray} 
Here, we have taken into account the term to order ${\cal O}(e^2 )$ only. 
Hence, what we need is to calculate $[\rlap{/}{P},M(P^2)]$, which
becomes in the gauge $[p_\mu , v_\mu ]=0$: 
\begin{equation} 
[ \rlap{/}{P},M(P^2 )]= -2ieQ F_{\mu\nu} \gamma_\nu p_\mu 
\frac{dM(p^2 )}{dp^2}+O(e^2, \partial^2 ) . 
\label{commutator1} 
\end{equation} 
Putting them together, we finally arrive at the expression: 
\begin{equation} 
-S_{kf} = \int \frac{d^4 p}{(2\pi )^4} 
\frac{M^2 (p^2 ) - p^2 \frac{d M^2 (p^2)}{dp^2}}{(p^2 
+ M^2 (p^2))^3} {\rm tr} \left(\frac{eQ}{2}\sigma\cdot F\right)^2 
(kf)\gamma_5 . 
\label{S_kf2} 
\end{equation} 
It was shown in the previous paper~\cite{Musakhanov:1996qf} 
that if we neglect the momentum dependence of the $M$, 
then we exactly reproduce the low-energy theorem in Eq.(\ref{theorem}). 
Hence, In order to prove that the gauged effective action exactly
satisfies the low-energy theorem of the axial anomaly, it is
enough to show that the following ratio 
$
R = \frac{J_1}{J_2} 
$
with 
\begin{eqnarray} 
J_1 = \int p^2 dp^2 
\frac{M^2 (p^2 ) - p^2 \frac{d M^2 (p^2)}{dp^2}}{(p^2 
+ M^2 (p^2))^3}, \,\,\,\,
J_2 =
\int dp^2 p^2 \frac{M^2(0)}{(p^2 +M^2(0) )^3} 
\end{eqnarray} 
becomes unity.  Replacing the variable~\cite{Plant:1997jr} by 
$ 
s\;=\; \frac{M^2 (p^2)}{p^2}, 
$
we immediately have the ratio $R=1$. 
As a result, the nonlocal chiral quark model from the instanton vacuum 
exactly satisfies the low-energy theorem given by Eq.(\ref{theorem}). 
 
Now, we briefly present some calculations related to Eq.(\ref{theorem1}). 
This matrix element can be written in the form: 
\begin{eqnarray} 
&& \langle 0| g^2 G\tilde G | g(\epsilon^{(1)} , q_1 ), g(\epsilon^{(2)} 
, q_2 )\rangle  \cr 
&=& \epsilon^{(1)a_1}_{\mu_1}  \epsilon^{(2)a_2}_{\mu_2} 
\int  \partial^{2}_{2} \,\partial^{2}_{1}\, \langle 0| 
T{g^2}G\tilde 
G A_{\mu_1}^{a_1}(x_1 ) A_{\mu_2}^{a_2}(x_2) 
|0 \rangle  \exp i(q_1 x_1 + q_2 x_2 ) dx_1 dx_2, 
\label{me2} 
\end{eqnarray} 
where $A_{\mu}^{a}(x )$ denotes a total gluon field.
$\epsilon^{(i)a_i}_{\mu_i}$ and $ q_i $ are the polarization vectors
and the momentum of gluons, respectively.    
 
As usual, we expand the total field $A_{\mu}^{a}(x )$ around the 
instanton background.  The main term in Eq.(\ref{me2}) is the 
contribution of the instanton background and is of order ${\cal O}(g^{-2})$. 
The next term is due to the perturbative fluctuations 
over the instanton background and corresponds to the contribution of 
order ${\cal O} (g^{2})$.  It is easy to see from previous 
considerations that the term to order $ {\cal O}(g^{-2})$ is given 
by the following expression: 
\begin{equation} 
Z_{N}^{-1} \int D\psi D\psi^\dagger \exp  (- S_{\rm eff} ) 
\left( Y_{G\tilde GAA +} (x) + Y_{G\tilde GAA -} (x)\right)
\label{GtildeGAAQ2} 
\end{equation} 
with 
\begin{eqnarray} 
Y_{G\tilde GAA \pm} &=& \pm \left(\frac{2V}{N}\right)^{N_f - 1} (i 
M)^{N_f} \int d^4 z\, f(x-z)  \cr 
&\times& \int dO 
(-\partial^{2}_{1})A_{\mu_1}^{I(\bar I)a_1}(x_1) 
(-\partial^{2}_{2})A_{\mu_2}^{I(\bar I)a_2}(x_2) 
 \det J_\pm (z) , 
\label{Y_GtildeGAAQ} 
\end{eqnarray} 
where the instanton(anti-instanton) is located at the point $z$ with 
its orientation $O$~\cite{Musakhanov:1996qf} over which we integrate. 
 
Repeating the bosonization, we obtain the result for the 
${\cal O} (g^{-2})$ contribution which is proportional to 
$
{\rm Tr} [ (i\rlap{/}{\partial} + iMF^2)^{-1}iM F^2 \gamma_5 ]. 
$
Hence, one can easily show that the contribution from the term to 
order ${\cal O}(g^{-2})$ vanishes. 
 
The contribution from the next order $({\cal O}(g^{2}))$ comes from 
two different diagrams.  The first diagram is the direct contribution of 
the operator $g^2 G\tilde G$ which is equal to 
$- g^2 G^{(1)} \tilde G^{(2)}$, where 
$ 
2G^{(1)} \tilde G^{(2)} = \epsilon^{\mu\nu\lambda\sigma} 
 G^{(1)a}_{\mu\nu} G^{(2)a}_{\lambda\sigma},\,\,\, 
G^{(i)a}_{\mu\nu} = 
\epsilon^{(i)}_{\mu}  q_{i\nu} - \epsilon^{(i)}_{\nu} q_{i\mu}. 
$
The factors at the vertices of the second one are 
$g \gamma_{\mu} \lambda_{a} /2$ and $i M f 
F^{2}\gamma_{5}N_{f}^{-1}$.  
 
A comparison with previous calculations ends up with the result that
the contribution from the second loop--diagram is equal in magnitude
but opposite in sign to that from the first one at $q^2 \rightarrow
0$.  Because of this cancellation, the contribution from order ${\cal
  O}(g^{2})$ vanishes in the limit $q^2 \rightarrow 0$.  Hence, we
arrive exactly at the same conclusion as in the previous case. 
 
From the above calculations, we conclude that the nonlocal chiral quark 
model from the instanton vacuum does exactly satisfy the low-energy 
theorems, once we have properly gauged the low-energy QCD partition 
function from the instanton vacuum. 
 
%\section{ Conclusion } 
{\bf 3.} The solution of the low-energy theorems of the axial anomaly 
is nontrivially related to the nonperturbative instanton 
structure of the QCD vacuum.  The effective action approach based on the 
instanton vacuum exactly satisfies these low-energy theorems. 
 
This approach provides a solid ground for future investigations into 
different amplitudes for the nonperturbative conversion of gluons 
into hadrons and photons.  Further studies relevant to all of 
these problems are under way. 
 
\section*{Acknowledgments} 
One of the authors(MM) acknowledges the warm hospitality of the 
Nuclear Theory Group of PNU, where this work was initiated.  The 
present work is supported by the Korean Research Foundation 
(KRF\--2002\--041\--C00067).  The work of MM is supported in part 
by INTAS, BMBF and SNF grants.

\end{document}